\documentclass[aps,pra,showpacs,twocolumn,amsmath,amssymb,superscriptaddress]{revtex4}

\usepackage[english]{babel}
\usepackage{latexsym}
\usepackage{graphics}
\usepackage{epsf} 
\usepackage{color}
\usepackage{url}
\usepackage[normalem]{ulem}
\usepackage{amsthm}
\usepackage{amsmath}

\def\be{\begin{equation}}
\def\ee{\end{equation}}
\def\bea{\begin{eqnarray}}
\def\eea{\end{eqnarray}}
\def\bi{\begin{itemize}}
\def\ei{\end{itemize}}
\def\bin{\begin{enumerate}}
\def\ein{\end{enumerate}}

\newcommand{\vect}[1]{\mathbf{#1}}

\usepackage{graphicx}

\begin{document}
\title{Simulation of non-Abelian lattice gauge fields with a single component gas
}


\author{Arkadiusz Kosior}
\affiliation{
Instytut Fizyki imienia Mariana Smoluchowskiego,
Uniwersytet Jagiello\'nski, ulica Reymonta 4, PL-30-059 Krak\'ow, Poland}

\author{Krzysztof Sacha}
\affiliation{
Instytut Fizyki imienia Mariana Smoluchowskiego,
Uniwersytet Jagiello\'nski, ulica Reymonta 4, PL-30-059 Krak\'ow, Poland}
\affiliation{
Mark Kac Complex Systems Research Center, 
Uniwersytet Jagiello\'nski, ulica Reymonta 4, PL-30-059 Krak\'ow, Poland}

\date{\today}

\begin{abstract}
We show that non-Abelian lattice gauge fields can be simulated with a single component ultra-cold atomic gas in an optical lattice potential. An optical lattice can be viewed as a Bravais lattice with a $N$-point basis. An atom located at different points of the basis can be considered as a {\it particle} in different internal states. The appropriate engineering of tunneling amplitudes of atoms in an optical lattice allows one to realize U$(N)$ gauge potentials and control a mass of {\it particles} that experience such non-Abelian gauge fields. We provide and analyze a concrete example of an optical lattice configuration that allows for simulation of a static U(2) gauge model with a constant Wilson loop and an adjustable mass of {\it particles}. In particular, we observe that the non-zero mass creates large conductive gaps in the energy spectrum, which could be important in the experimental detection of the transverse Hall conductivity. 
\end{abstract}


\maketitle

\section{Introduction}

Ultra-cold dilute atomic gases are ideal laboratories for the realization of quantum emulators \cite{Feynman1982}, i.e. for the creation of controllable devices that can simulate other systems of interest. At low temperatures atoms behave like point particles but their internal structure allows for the flexible manipulation of the effective particle interactions and the preparation of external potentials whose shape can be chosen nearly at will \cite{exptech}. Atoms are charge neutral and seem not suitable to simulate the orbital magnetism. However, it is possible to create artificial magnetic fields, i.e. to prepare specific conditions where the motion of neutral particles mimics the dynamics of charged particles in an effective magnetic field \cite{Dalibard2011,goldman2013}. Owning to the internal structure of atoms it is also possible to simulate non-Abelian gauge fields \cite{Lin2011,Wang2012,Cheuk2012,Zhang2012,Fu2013,Zhang2013,Qu2013,LeBlanc2013}. They emerge when the centre of mass motion of an atom 
is coupled to internal atomic degrees of freedom. 

Atomic gases in optical lattice potentials can simulate lattice gauge fields. The latter arise, e.g., in quantum electrodynamics and quantum chromodynamics when configuration space is discretized \cite{lattice_gauge}. Lattice models describe also crystalline materials in condensed-matter physics. Quantum emulators with ultra-cold atomic gases open possibilities for deep understanding of fundamental problems like quark confinement, high-temperature superconductivity or strongly interacting counterparts of topological insulators \cite{goldman2013}. 

Although the ultimate goal is to realize a quantum simulator of dynamical non-Abelian fields, the first step is to implement static field configurations. There are various proposal for the creation of non-Abelian static gauge potentials for atomic gases in optical lattices \cite{Osterloh2005,kubasiak,Goldman2009,Mazza2012,struck2012a,Goldman2013a,Tagliacozzo2013}. Some of them involve lattice shaking or light assistant tunneling but all of them try to couple the centre of mass motion of an atom with an atomic internal structure, { i.e. they employ multi-component gases \cite{Dalibard2011,goldman2013} (see also the recent proposal for the realization of a spin-orbit coupling via photon-assistant band hybridization \cite{Zhang2014}).  In the present Letter we show that the centre of mass degree of freedom is sufficient to model static non-Abelian gauge fields  if tunneling amplitudes in an optical lattice are engineered appropriately. In the paper we focus on the model with a constant Wilson loop over a square 
plaquette. 

\section{Mimicking multicomponent particles with an optical lattice}

A single component atomic gas in an optical lattice can be described by the Hamiltonian $H=-\sum_{i,j} ( J_{ij}\hat{\psi}_{i}^{\dagger}{\hat{\psi}_{j}}+h.c.)$, where $\hat\psi_i$ is the bosonic or fermionic field operator that annihilates an atom at $i$-site of the lattice and $J_{ij}$'s stand for tunneling amplitudes between the sites \cite{exptech}. We omit atom interactions and assume that tunneling of an atom from a given site is not restricted to the nearest neighbors. The lattice geometry can be arbitrary, i.e. it can form a simple Bravais lattice or a crystal structure with a $N$-point basis. An example of a two-dimensional (2D) squared Bravais lattice with a two-point basis is shown in Fig.~\ref{dimers}(a). Let us identify a position of an atom in different points of the basis with different internal states of a {\it particle} located at a given site of the Bravais lattice. That is, we define $N$-component operators $\hat\Psi_i$ built of the $\hat\psi_j$ operators that correspond to different points 
of the 
basis of the same Bravais lattice sites. Then, we ask the question if the Hamiltonian of a system can be written in the form
\be
\label{hamiltonian}
H= - \sum_{i,j} \left( t_{ij}\, \hat{\Psi}_{i}^{\dagger} U_{ij}{\hat{\Psi}_{j}}  +h.c. \right),
\ee
where real $t_{ij}\ge0$ and the tunnelings of a {\it particle} between the Bravais lattice sites is described by unitary link operators $U_{ij}$. It turns out it is possible and the Hamiltonian (\ref{hamiltonian}) can simulate non-Abelian gauge fields provided the tunneling amplitudes of atoms, $J_{ij}$, fulfill the specific conditions. 

In our approach, an optical lattice structure itself induces non-Abelian behaviour. The multicomponent field operators $\hat\Psi_i$ can be defined if one creates an optical lattice corresponding to a Bravais lattice with a basis like, e.g., in Fig.~\ref{dimers}(a). However, a {\it particle} described by $\hat\Psi_i$ does not necessary need to be spatially well separated from the others, and easily identifiable in the lattice. {\it Particles} may be assigned arbitrary at one's convenience even if an optical lattice is a simple Bravais lattice, see Fig.~\ref{dimers}(b).  The main difficulty is to ensure the unitarity of the transition matrices $U_{ij}$ what can be overcome due to the great flexibility and sophistication in optical lattice engineering \cite{exptech}. 
Our method is flexible in the manipulation of particle interactions because it does not involve explicitly internal atomic structure. For example, in order to simulate the integer quantum Hall effect there is no need to turn off the particle interactions because a single component Fermi gas does not interact at low temperatures. The method allows one to easily incorporate a {\it particle's} mass term in a lattice gauge Hamiltonian \cite{lattice_gauge}. This can be done by introducing the tunnelings between lattice sites corresponding to the different components of $\hat\Psi_i$. It results in an extra term $-\sum_{i} \hat{\Psi}_{i}^{\dagger} \mathcal{M} {\hat{\Psi}_{i}}$ in (\ref{hamiltonian}), where $\mathcal{M}$ is a matrix  that describes masses of the {\it particle} components.

\begin{figure}[bt]
\begin{center}
\begin{minipage}{0.45\columnwidth}
 \resizebox{0.9\columnwidth}{!}{\includegraphics{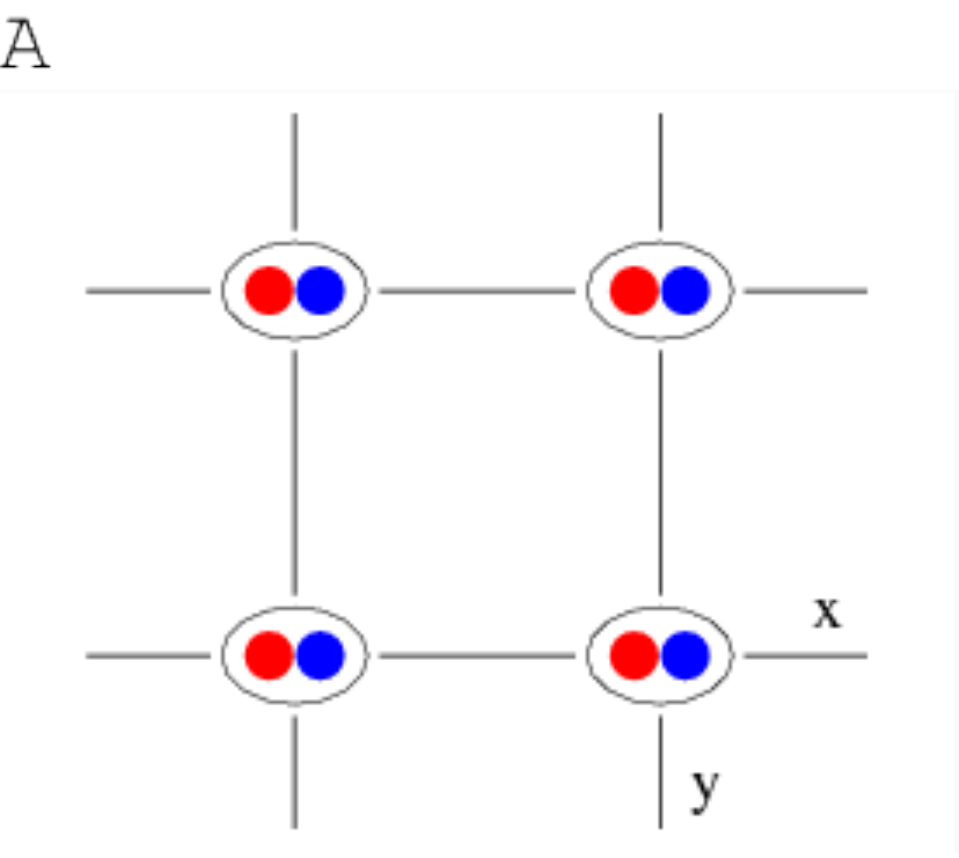}} 
 \resizebox{0.9\columnwidth}{!}{\includegraphics{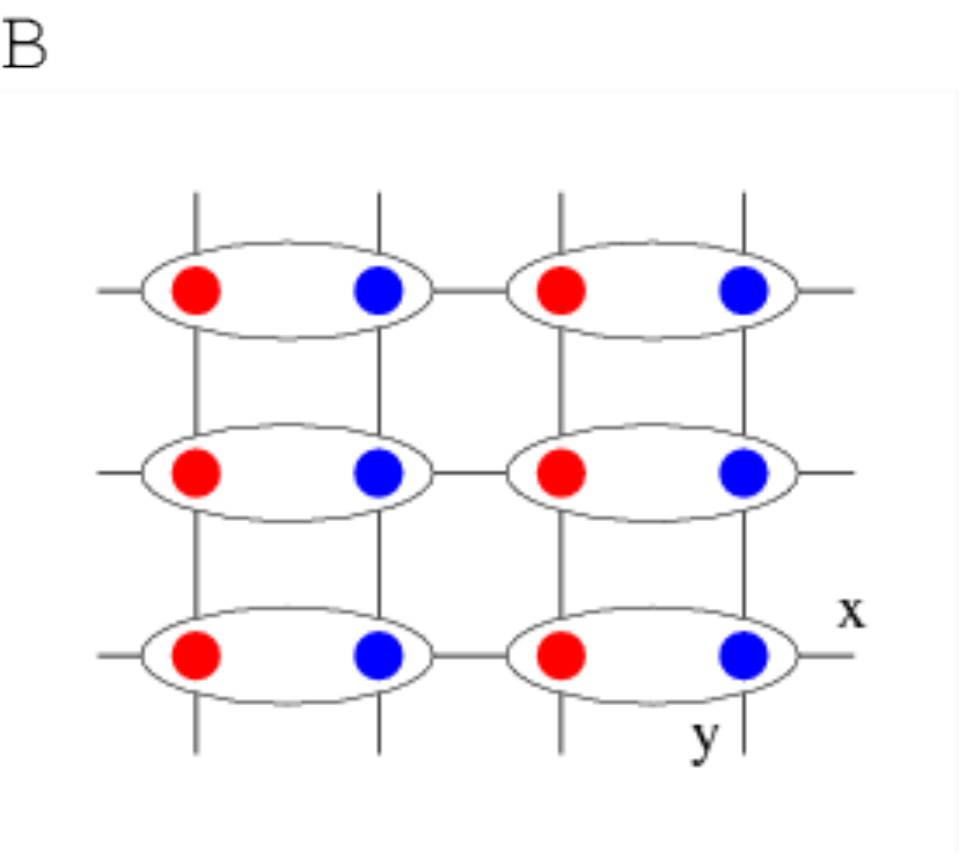}}
\end{minipage}
\begin{minipage}{0.45\columnwidth}
 \resizebox{0.8\columnwidth}{!}{\includegraphics{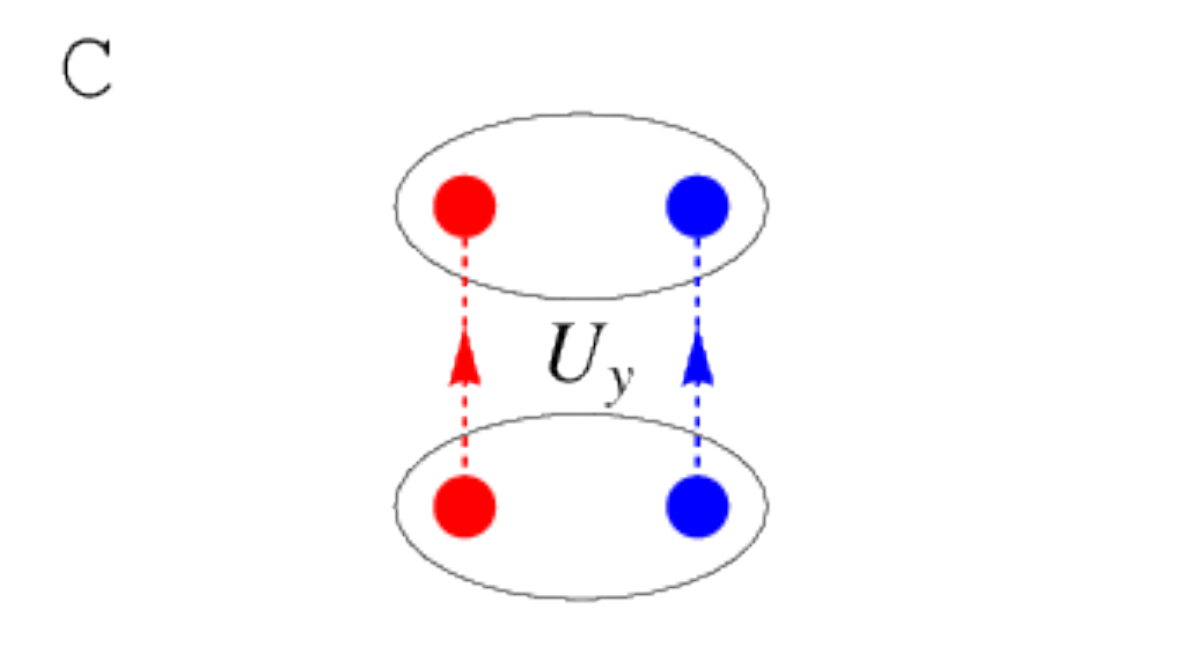}}
  \resizebox{0.8\columnwidth}{!}{\includegraphics{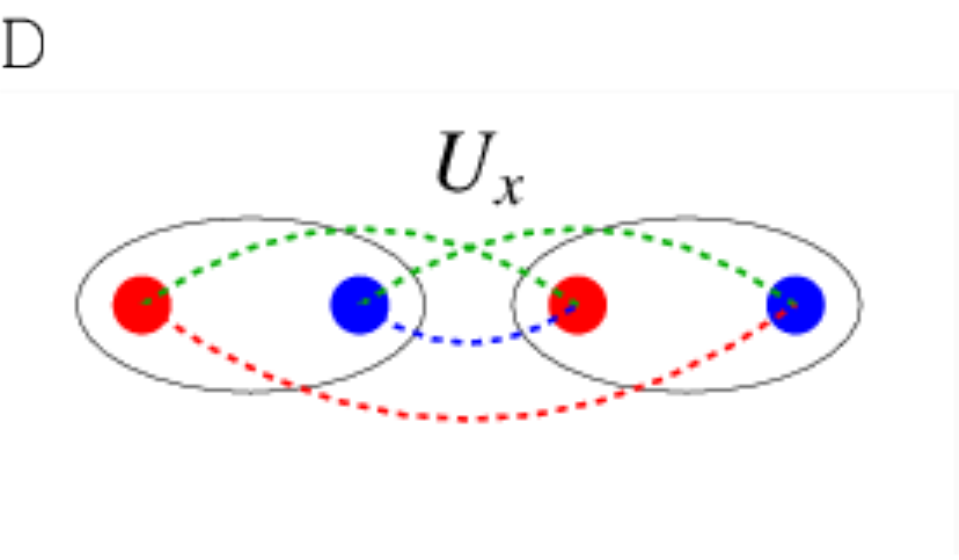}}
 \resizebox{0.8\columnwidth}{!}{\includegraphics{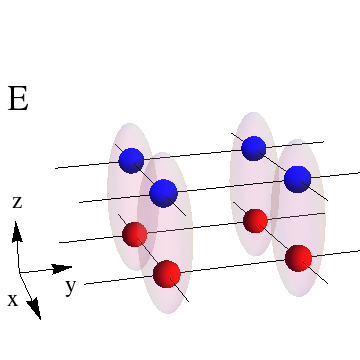}}
\end{minipage}
\end{center}
\caption
{
(color online) The key idea of our approach is that non-Abelian lattice gauge fields can emerge from Abelian fields, provided adequate hoppings of atoms in an optical lattice are imposed. For example, the two-point basis associated with the 2D crystal structure in (a) can be assigned with the two-component {\it particles} (let us call them {\it dimers}). It turns out that the spatial separation of the {\it dimers} is irrelevant, and therefore the components can be chosen arbitrary even on a simple Bravais lattice (b) if the hoppings of the {\it dimers} are described by unitary matrices. To ensure the unitarity we have to engineer the adequate tunnelings of atoms (c)-(d). The latter can be more easily realized experimentally if the {\it dimers} are oriented perpendicularly with respect to the Bravais lattice (e).
}
\label{dimers}
\end{figure}

\begin{figure}[tb]
\begin{center}
\resizebox{0.45\columnwidth}{!}{\includegraphics{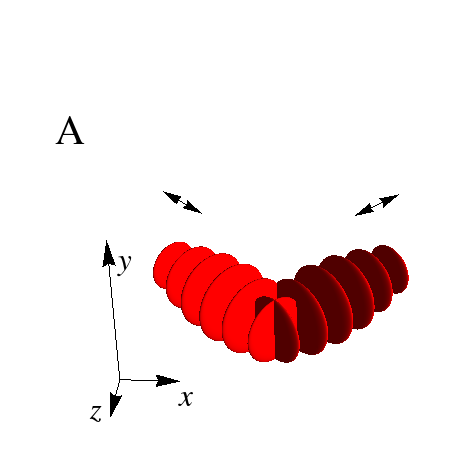}}
\resizebox{0.45\columnwidth}{!}{\includegraphics{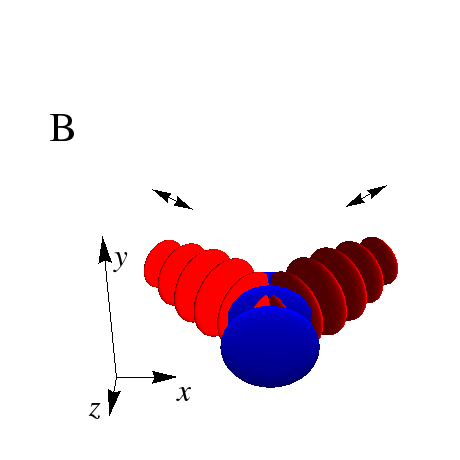}}
\resizebox{0.49\columnwidth}{!}{\includegraphics{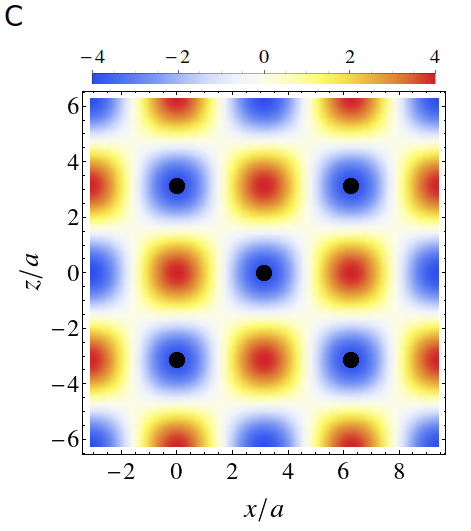}}
\resizebox{0.49\columnwidth}{!}{\includegraphics{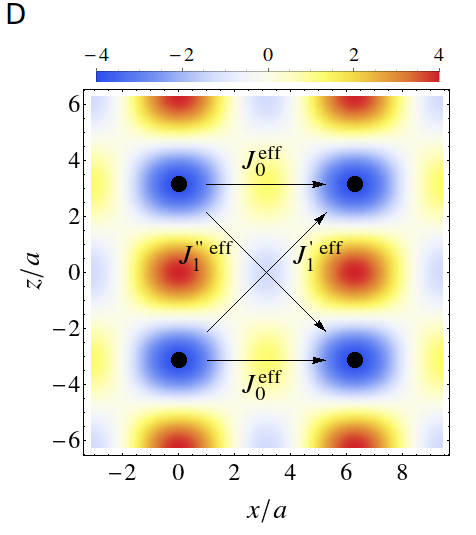}}
\caption{
(color online) The procedure of engineering of the optical lattice potential which allows for the realization of a non-Abelian U(2) gauge field. First, we superpose two orthogonal standing waves along the $\hat{x}\pm\hat{z}$ directions (a), which give rise to the checkerboard potential (c). The two standing waves can be independently shaken by introducing periodic modulations of the phases of the superimposed waves. Then, we add an extra standing wave along the $z$ direction (b), which breaks the separability of the potential along the $\hat{x}\pm\hat{z}$ directions and changes the potential to the square anisotropic lattice (d). Controlling the shaking parameters allows us to control the tunneling amplitudes of atoms which are indicated in (d). Finally, an additional standing wave along the $y$ direction and a trapping potential $V_{trap}(z)$ are added. The latter selects out the pairs of lattice sites that leads to the creation of the crystal structure with the 2D squared Bravais lattice and the two-point 
basis as in Fig.~\ref{dimers}(
e). The black dots indicate the positions of the potentials' minima.}

\label{realizacja}
\end{center}
\end{figure}

\section{Experimental realization}

As a concrete example let us show how to realize experimentally the U(2) non-Abelian gauge field on the 2D squared Bravais lattice with $H=H_0-\sum_{m,n} \hat{\Psi}_{m,n}^{\dagger} \mathcal{M} {\hat{\Psi}_{m,n}}$ where
\be
\label{hamiltonian_spec}
H_0= - \sum_{m,n} \left( t_x\hat{\Psi}_{m+1,n}^{\dagger} U_x {\hat{\Psi}_{m,n}}  + t_y\hat{\Psi}_{m,n+1}^{\dagger}  U_y {\hat{\Psi}_{m,n}} +h.c. \right),
\ee
and $\mathcal{M}=m_{\rm eff}\sigma_1$. The mass term can be also interpreted as the presence of a Zeeman field experienced by a {\it particle} with a psuedospin.
The Bravais lattice points are $\vect{r}_{m,n}=a(m\hat{x}+n\hat{y})$ and the link operators $U_{x}=e^{iA_{x}}$ and $U_{y}(m,n)=e^{iA_{y}}$ with the matrix valued vector potential
\bea
A_x=\alpha \sigma_2, \quad A_y(m,n)=\beta \sigma_3-2\pi(m\phi_x+n\phi_y),
\label{vect_pot}
\eea
where $\sigma_i$ are the Pauli matrices and $\alpha$, $\beta$, $\phi_x$ and $\phi_y$ are real parameters. The two-dimensional optical lattices presented in Fig.~\ref{dimers}(a)-(b) are examples which allow for the realization of such a gauge field. In Fig.~\ref{dimers}(c)-(d) there are indicated hoppings whose tunneling amplitudes must be controlled. This control seems to be easier experimentally if we consider the 2D Bravais lattice with the basis whose points are displaced along the $z$ direction, e.g., by the Bravais lattice constant $a$, see Fig.~\ref{dimers}(e). The different points of the basis, $\vect{R}_{m,n,\pm} =\vect{r}_{m,n}\pm \frac{a}{2}\hat{z}$, are identified with the internal states of a {\it particle} located at $\vect{r}_{m,n}$ which we will call a {\it dimer}. 

To start with, we construct a 3D lattice potential $V$, out of four non-interfering standing waves with an additional trapping potential, $V_{trap}(z)$, that selects out the pairs of sites corresponding to the {\it dimer} components, i.e. $V=V_1(x,z) +V_2(y)+V_{trap}(z)$, where

\begin{gather}
 V_1 = V_0 \left\{\cos\left[k(x+z)\right] + \cos\left[k(x-z)\right]  
+ \Lambda \cos\left(k z\right)\right\}, 
\label{lattice_potential}
\end{gather}
and $V_2=-\Gamma \cos\left(k y\right)$ with $k=2 \pi/a$ and $V_0 $, $\Lambda$, $\Gamma>0$, see Fig.~\ref{realizacja}. The potential $V$ does not couple the $y$ and $z$ variables and consequently there is no mixing of the {\it dimer} components when the {\it dimer} tunnels along the $y$ direction, i.e. the link operator $U_y$ is diagonal as in (\ref{vect_pot}). Furthermore, the complex phases in $U_y$ can be realized using laser-assisted tunneling \cite{goldman2013,Kolovsky2011, Aidelsburger2011,Aidelsburger2013, Miyake2013}. The laser-assisted tunneling technique requires a potential energy gradient provided by the gravity or magnetic field gradients, and two running-wave beams with a difference $\delta \vect{k}$ of the wave-vectors. The frequency difference $\omega$ of the running beams matches the energy difference, between the neighboring sites, corresponding to the potential energy gradient. It results in an extra term $\hbar \omega y/a +\Omega \sin (\delta \vect{k} \cdot \vect{r} -\omega t)$ in the potential $V$.  The phase an atom acquires when it tunnels along the $y$ direction is position dependent, i.e. it equals $\delta \vect{k} \cdot \vect{R}_{m,n,\pm}= \pm \beta-2\pi(m\phi_x +n \phi_y)$ where $\vect{R}_{m,n,\pm}$ is the position of an atom.  

Finally, in order to control the transitions in accordance with $U_x$ we invoke the periodic lattice shaking \cite{eckardt05,struck2011,targonska2012,kosior2013}. The lattice shaking demands an additional term, $-[K_1 (x+z)/a +K_2 (z-x)/a] \sin \omega t$, in the potential $V$ where $\omega$ is the shaking frequency (which we assume the same as the frequency difference of the running beams in the laser assisted tunneling technique) and $K_1$, $K_2$ are the shaking strengths. If $\hbar \omega$ is greater than the bare atomic hopping amplitudes but smaller than the energy gap between the lowest and first excited energy bands of the optical lattice, the evolution can be time averaged \cite{Holthaus_2012}. The averaging procedure comes down to the renormalization of the atomic hopping amplitudes. For simplicity we assume that $K_1$, $K_2$ and their difference are much larger than $\Omega$. Then, the influence of the running waves on the renormalization of the tunneling amplitudes along the $x$ and $z$ directions is very small and can be neglected. Due to the symmetries of $V$, there are only two parameters that describe the hoppings of atoms along the $x$ direction, i.e. $\vect{R}_{m,n,\pm}\rightarrow\vect{R}_{m+1,n,\pm}$ are described by the bare amplitude which we denote by $J_0$ and $\vect{R}_{m,n,\pm}\rightarrow\vect{R}_{m+1,n,\mp}$ by $J_1$, cf. Fig.~\ref{realizacja}. The corresponding terms of the 
Hamiltonian can be represented as $t_x\hat{\Psi}_{m+1,n}^{\dagger} U_x \hat{\Psi}_{m,n}$ where $t_xU_x=J_0+J_1\sigma_1$ and $\hat\Psi_{m,n}^T=[\hat\psi_{(m,n,+)},\, \hat\psi_{(m,n,-)}]$. The shaking partially breaks that symmetry and hence the effective hoppings are
\be
t_xU_x=
\left( 
\begin{array}{cc}
J_{0}^{\mbox{{\tiny eff}}}  & \quad J_{1}^{'\mbox{{\tiny eff}}}  \\
J_{1}^{''\mbox{{\tiny eff}}}   &  \quad J_{0}^{\mbox{{\tiny eff}}}
\end{array}
\right),
\ee
with $J_{1}^{'\mbox{{\tiny eff}}}=J_1\mathcal{J}_0 \left(2 K_1/\hbar \omega\right)$,
$J_{1}^{''\mbox{{\tiny eff}}}=J_1\mathcal{J}_0 \left(2 K_2/\hbar \omega\right)$ and
$J_{0}^{\mbox{{\tiny eff}}}=J_0\mathcal{J}_0 \left((K_1-K_2)/\hbar \omega\right)$ where $\mathcal{J}_0$ stands for the Bessel function \cite{Holthaus_2012}. Our goal to achieve the unitarity of the $U_x$ operator can be reached by a proper choice of $K_1$, $K_2$ and $\omega$. If we set $J_{1}^{'\mbox{{\tiny eff}}}=-J_{1}^{''\mbox{{\tiny eff}}}$ and define $t_x=\sqrt{(J_0^{\mbox{{\tiny eff}}})^2 +(J_{1}^{'\mbox{{\tiny eff}}})^2}$, then $U_x=e^{iA_x}$ where $A_x$ is given in (\ref{vect_pot}) with $\tan\alpha=-J_{1}^{'\mbox{{\tiny eff}}}/J_0^{'\mbox{{\tiny eff}}}$. The shaking of the lattice allows also for the controlling of effective masses of the {\it dimer} components which are represented by the matrix $\mathcal{M}=m_{\rm eff}\sigma_1$.  Indeed,  $m^{\mbox{{\tiny eff}}}=J_2\mathcal{J}_0((K_1+K_2)/\hbar \omega)$ where $J_2$ is the bare amplitude of atom tunneling between $\vect{R}_{m,n,\pm}$ and $\vect{R}_{m,n,\mp}$.

 Finally let us consider examples of experimentally attainable parameters. The bare tunneling amplitudes can be calculated by Fourier transform of the dispersion relation of the lowest energy band of an atom in the potential $V_1(x,z)+V_2(y)$.
For instance, setting $V_0=6 E_R$, $\Lambda=1$, where $E_R$ is the recoil energy, we get  $J_0 \approx 0.035E_R $,  $J_1 \approx 0.015E_R $ and $J_2 \approx 2.2\cdot10^{-3}E_R$.   In our model there are two angles $\alpha$ and $\beta$ which determine the non-Abelian behaviour, see (\ref{vect_pot}). The angle $\beta $ can be set nearly arbitrary by a suitable choice of the direction of the vector $\delta \vect{k}$ while the value of $\alpha$ depends on the shaking parameters. For example, for $K_1/ \hbar \omega \approx 1.38 $ and $K_2/ \hbar \omega \approx 7.01 $ one gets  $t_x \approx 1.3\cdot 10^{-3}E_R$, $\tan \alpha \approx 0.28$ and $m\mbox{{\tiny eff}} \approx 10^{-3}E_R$. Then, and for $\beta=\pi/2 $, the Wilson loop $|W| \approx 1.7$, see Sec.~\ref{thc}. In order to get $|W|$ very close to zero,  we need $\alpha = \pi/4$ that can be achieved, for example, for $K_1/\hbar \omega \approx 7.81$, $K_2/\hbar \omega \approx 10.2$. The effective tunneling along the $y$ direction
\be
 J_y^{\mbox{\tiny eff}}=J_y\mathcal{J}_1\left[\frac{2\Omega}{\hbar \omega} \sin(\pi \phi_y)\right] e^{i [ 2\pi(\phi_x m +\phi_y n) \pm \beta]},
\ee
depends on the amplitude $\Omega$ of the running beams and on the bare tunneling amplitude $J_y$. For  $\Omega/\hbar\omega=0.1$, $\Gamma=6 E_R$ and $\phi_y=1/2$ we obtain  $t_y=|J_y^{\mbox{\tiny eff}}| \approx 2.5\cdot 10^{-3}E_R$. Note that the assumed condition $\Omega\ll K_{1,2},|K_1-K_2|$ is also fulfilled.}


\begin{figure}[htb]
\begin{center}
\resizebox{0.7\columnwidth}{!}{\includegraphics{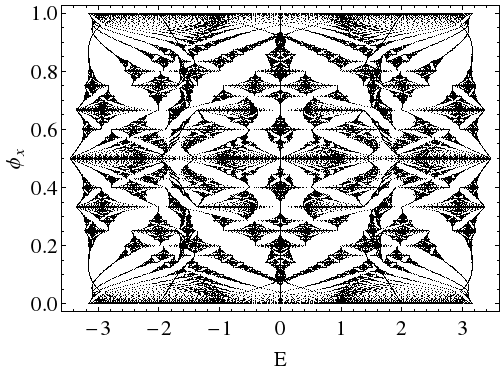}}
\resizebox{0.7\columnwidth}{!}{\includegraphics{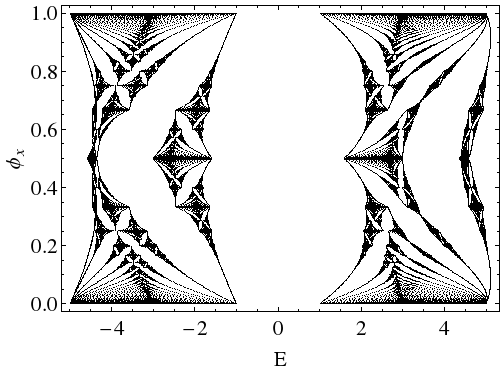}}
\resizebox{0.7\columnwidth}{!}{\includegraphics{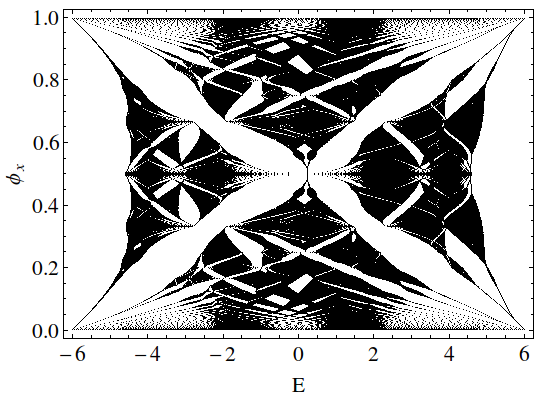}}
\caption{
The energy spectrum $E(\phi_x)$ as a function of the Abelian magnetic flux $\phi_x$ for $t_y=t_x=1$, $\alpha=\beta=\pi/3$ and the mass parameter $m_{\rm eff}=0$ (top panel) and $m_{\rm eff}=3$  (middle panel). The energy spectrum at the bottom panel is calculated for  $t_y/t_x=2$, $\alpha=0.3$, $\beta=\pi/3$ and  $m_{\rm eff}/t_x= 0.8$, which correspond to the experimental parameters discussed in the main text. When the mass of the components is introduced and increases, the reflection symmetry of the spectrum is broken and the spectrum widens and eventually grows apart. 
}
\label{motyle}
\end{center}
\end{figure}

\section{Transverse Hall conductivity}
\label{thc}

Let us switch to an analysis of the model described by the Hamiltonian (\ref{hamiltonian_spec}) supplemented with the mass term. In the Letter we focus on the effects induced by the non-zero masses of the components, whereas the detailed analysis of massless non-Abelian model can be found in Ref.~\cite{Burello2013}.  In the following we assume $t_x=t_y=1$. The Hamiltonian is invariant under any local U(2) gauge transformation, i.e. $\hat{\Psi}^{'}_{m,n}=T_{m,n}\hat{\Psi}_{m,n}$, $U^{'}_{x}=T_{m+1,n}U_{x} T_{m,n}^{\dagger}$ and similarly for $U_y$, where $T_{m,n}\in \mathrm{U}(2)$. In order to distinguish between the different classes of gauge fields it is useful to define the Wilson loop over a square oriented  plaquette of neighboring sites, $W= \mbox{tr}U_xU_yU_x^{\dagger}U_y^{\dagger}$, which is a gauge invariant quantity \cite{kubasiak}. Although a single link operator, $U_x$ or $U_y$, can always be transformed to the identity by the gauge transformation, a product of link operators over a closed loop 
cannot be trivialized when a gauge field is genuinely non-Abelian. Hence, our system is non-Abelian if and only if $|W|=2|\cos ^2 \alpha +\cos(2\beta) \sin^2 \alpha|\ne2$. It is 
known that the condition $\left[U_x,U_y\right] \ne 0 $ is not sufficient to reach the non-Abelian regime \cite{kubasiak,goldman2013}. However, if $U_x$ and $U_y$ do not commute and anti-commute, $\left\{U_x,U_y\right\} \ne 0$, one can show that the gauge theory is genuinely non-Abelian, see Appendix A for a general U$(N)$ case. 

 Although the presence of the Abelian phase $\phi_x$ in (\ref{vect_pot}), does not modify the modulus of the Wilson loop, it induces a uniform Abelian magnetic field $\vect{B}=-2\pi\phi_x\hat z$ and breaks the spatial periodicity of the system. The other Abelian phase $\phi_y$ can be eliminated by the gauge transformation and therefore we set $\phi_y=0$. The periodicity is restored when the phase $\phi_x$ is a rational number, i.e. $\phi_x = p/q$ with $p$, $q$ being integers. Then, the Bloch theorem guarantees that solutions of the time-independent Schro\"{o}dinger equation are of the form $\Phi_{\vect{k}}(m,n) = e^{i k_x m} e^{i k_y n}\vect{u}_{\vect{k}}(m)$, where $\vect{u}^T_{\vect{k}}(m)=[u_{\vect{k}}(m,+),u_{\vect{k}}(m,-)]$ is a periodic function of $m$ \cite{Burello2013}. After the substitution the Schro\"{o}dinger  equation reduces to a generalized Harper equation and therefore we should expect a spectrum of the Hofstadter butterfly type \cite{kubasiak,Dalibard2011,goldman2013}. Indeed, the 
characteristic fractal structure of 
the energy levels plotted against the Abelian flux $\phi_x$ for the mass term $\mathcal{M}=m_{\rm eff}\sigma_1=0$ is shown in Fig.~\ref{motyle}.  The fractal structure depends on the non-Abelian phases $\alpha$ and $\beta$. What is more, the spectrum gets modified when the {\it particle} components acquire mass. Firstly, the reflection symmetry of the spectrum with respect to $E=0$ is broken. Secondly, with increasing $m_{\rm eff}$, large energy gaps are formed and the spectrum starts to spread and eventually separates into two parts, see Fig.~\ref{motyle}. 

\begin{figure}[tb]
\begin{center}
\resizebox{0.7\columnwidth}{!}{\includegraphics{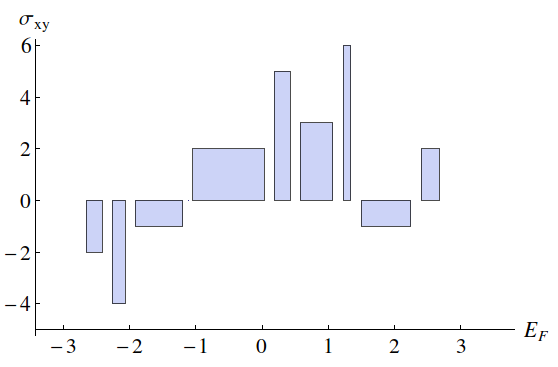}}
\resizebox{0.7\columnwidth}{!}{\includegraphics{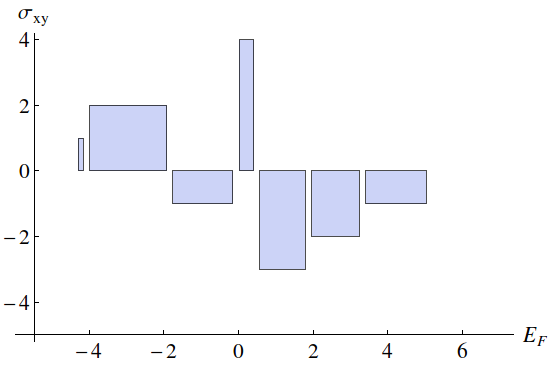}}
\caption{
Transverse Hall conductivity $\sigma_{xy}$ (in units of $1/h$) as a function of the Fermi energy for $t_y=t_x=1$, $\alpha=\beta=\pi/3$, $m_{\rm eff}=2$, $\phi_x=2/5$ (top panel) and  e $t_y/t_x=2$, $\alpha=0.3$, $\beta=\pi/5$ ,  $m_{\rm eff}/t_x= 0.8$, $\phi_x=1/4$ (bottom panel).
}
\label{conductivity}
\end{center}
\end{figure}

A topological insulator is characterized by a bulk energy gap like ordinary insulators. However, the finite system possesses conducting states which are gapless states on the edge and which are deeply related to the topology of bulk states \cite{Hasan2010,Qi2011,goldman2013}. Indeed, the topological invariants, Chern numbers, determine the transverse Hall conductivity. Let us consider the ground state of non-interacting fermions corresponding to the Fermi energy $E_F$.  We denote with $E_{\mathcal{N}}(\vect{k})$ a $\mathcal{N}$-th energy band and $|\mathcal{N}_{\vect{k}}\rangle$ a non-degenerate eigenstate of $H|\mathcal{N}_{\vect{k}}\rangle = E_\mathcal{N}(\vect{k}) |\mathcal{N}_{\vect{k}}\rangle$. The energy bands $E_\mathcal{N}(\vect{k})$ are smooth functions in the Brillouin zone. The latter forms a 2D torus $\mathrm{T}^2$. With each band one can assign the geometrical non-Abelian Berry curvature
\be
F_n = \left(\partial_{k_x} \mathcal{B}_y - \partial_{k_y}\mathcal{B}_x +[\mathcal{B}_x,\mathcal{B}_y] \right)dk_xdk_y,
\label{curvature}
\ee
where $\mathcal{B}_{\mu}  = \langle \mathcal{N}_{\vect{k}}|\partial_{k_{\mu}}|\mathcal{N}_{\vect{k}}\rangle$ is the Berry connection \cite{Hasan2010,Qi2011,goldman2013,goldman2007}. The Chern numbers, $c_n = \frac{1}{2\pi i} \int_{\mathrm{T}^2} \mbox{tr}F_n$, are integer-valued topological invariants which fully determine the transverse Hall conductivity, $\sigma_{xy}  = -\frac{1}{h} \sum_n c_n$, where the summation runs over all the $\mathcal{N}$ energy bands below the Fermi energy $E_F$ \cite{goldman2007}.  In Fig.~\ref{conductivity} we present examples of $\sigma_{xy}(E_F)$. Importantly, the non-zero mass is responsible for the formation of large energy gaps corresponding to $\sigma_{xy}\ne 0$, which can be very important in the experimental detection of the quantization of the Hall conductivity. Different proposals of experimental detection of the Hall conductivity and topological invariants in cold atom systems are currently under investigation, see \cite{goldman2013} and the 
references therein.

\section{Conclusions}

Summarizing, we have shown that a single component ultra-cold atomic gas in an appropriately engineered optical lattice potential can simulate non-Abelian lattice gauge fields. An optical lattice can be identified with a crystal structure, i.e. a Bravais lattice with a $N$-point basis. We identify the different points of the basis with the internal states of a {\it particle} that can move in the Bravais lattice. It can be done if the hoppings of the {\it particle} are described by the unitary operators. A concrete proposal of the experimental realization of a non-Abelian SU(2) gauge field model is given and analyzed. Our idea is a novel route to the realization of quantum emulators of non-Abelian gauge fields in ultra-cold atomic gases. It has advantages over more common proposals where internal atomic states are coupled to the centre of mass motion of an atom. For example, the manipulation of the particle interactions is more flexible because the internal structure of an atom is not explicitly involved 
in our proposal. Finally, the 
massive components greatly modify the Hofstadter-like energy spectrum, creating the large energy gaps with the non-zero transverse Hall conductivity.  

 We have considered mapping between a system of a spinless particle in an optical lattice potential and a system of a {\it particle} with a pseudo-spin degree of freedom but in a simpler lattice. If such a mapping is possible, a theoretical description of the translational motion of the spinless particle possesses simpler interpretation.

\section{Acknowledgements}

We acknowledge support of Polish National Science Centre via project DEC-2012/04/A/ST2/00088. AK acknowledges support in a form of a special scholarship of Marian Smoluchowski Scientiﬁc Consortium
“Matter Energy Future” from KNOW funding.

\section*{Appendix A}


\begin{figure}[htb]
\begin{center}
\resizebox{0.5\columnwidth}{!}{\includegraphics{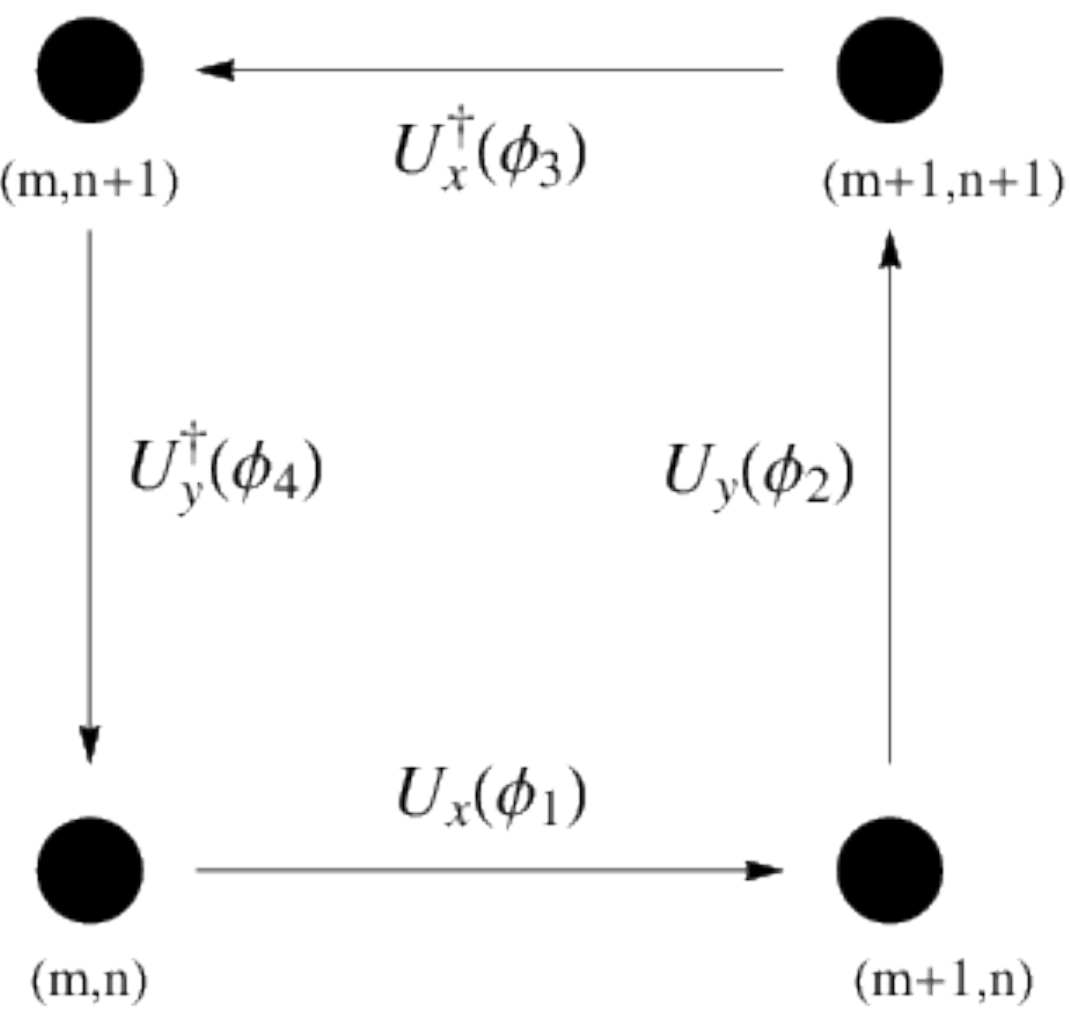}}
\caption{
 Square unit plaquette of a $\mathrm{U(N)}$ lattice with the constant Wilson loop. 
}
\label{plaquette}
\end{center}
\end{figure}

In this section we justify the necessary and sufficient conditions to reach the non-Abelian $\mathrm{U(N)}$ regime. The proof covers the case of the constant Wilson loop on the square lattice, $W= \mbox{tr}\;U_\square$, where $U_\square$ is a product of link operators around a unit plaquette, $ U_\square=U_x(\phi_1) U_y(\phi_2) U_x^{\dagger}(\phi_3) U_y^{\dagger}(\phi_4) $, where the operators $ U_{\mu}(\phi_j) = e^{i\phi_j} U_{\mu}$ with $U_{\mu} \in \mathrm{SU(N)}$ and $\phi_j$ being an arbitrary phase, see Fig. \ref{plaquette}. If the operator $U_\square$ can not be gauge transformed into $U_\square=e^{i \Psi} \mathrm{1}$ (which is the case if $|W|\ne N$ \cite{kubasiak,goldman2013}), then the lattice is genuinely non-Abelian.  On the other hand if $U_\square=e^{i \Psi} \mathrm{1}$, then the system is a collection of independent Abelian systems subjected to a uniform flux $\Psi$ per plaquette, cf. \cite{Barnett2012}.

\emph{Lemma.}  In order to achieve the non-Abelian $\mathrm{U(N)}$ regime it suffices to fulfill the condition $ U_x U_y \ne (\pm1)^{(N+1)} U_y U_x $.
\begin{proof}
 The product of the link operators in the Abelian case must be proportional to the identity
\be
\tag{A.1}
U_\square=U_x U_y U_x^{\dagger}U_y^{\dagger} e^{i \Phi} = e^{i \Psi} \mathrm{1},
\label{UN_product}
\ee
where $\Psi$ is an arbitrary phase and  $ \Phi =\phi_1+\phi_2-\phi_3-\phi_4 $
is the total flux around a unit plaquette. After calulating the determinant one finds out that in fact $\Psi= \Phi +2\pi n/N, \,n\in\mathbb{Z}$ and so
\be
\tag{A.2}
U_x U_y = e^{2\pi i  n/N} U_y U_x.
\label{UxUy}
\ee
On the other hand, an arbitrary element $U$ of $\mathrm{SU(N)}$ group can be expressed in terms of its  generators, i.e. the constituents of the  Lie algebra $\mathfrak{su}\mathrm{(N)}$:  $ U =  a_0 \, \mathrm{1} + i \sum_k a_k \, \rho_k$, $ a_0^2 +\sum_k a_k^2 =1$, where $a_0, a_k \in \mathbb{R}$ and $\rho_k \in \mathfrak{su}\mathrm{(N)}$. Since the elements of $\mathfrak{su}\mathrm{(N)}$  are traceless, then it follows that the phase factor in (\ref{UxUy}) is real, which proofs that the condition for the Abelian case, Eq.~(\ref{UN_product}), is equivalent to  $ U_x U_y = (\pm1)^{(N+1)} U_y U_x$.

\end{proof}



\end{document}